\documentclass[aps,notitlepage,a4paper,superscriptaddress,11pt]{article}

\usepackage[utf8]{inputenc}
\usepackage[dvips]{color}
\usepackage[toc]{appendix}
\usepackage[hang,flushmargin]{footmisc} 
\usepackage{titlesec,titletoc,ntheorem-hyper,authblk,abstract,latexsym,graphicx,microtype,booktabs,amsmath,fancyhdr,wrapfig,array,amsfonts, amssymb,geometry,appendix,cite,caption,subcaption}
\usepackage{float}
\usepackage{hyperref,cleveref}

\newcommand{\be}{\begin{equation}}
\newcommand{\ee}{\end{equation}}
\newcommand{\bea}{\begin{eqnarray}}
\newcommand{\eea}{\end{eqnarray}}
\newcommand\blfootnote[1]{%
  \begingroup
  \renewcommand\thefootnote{}\footnote{#1}%
  \addtocounter{footnote}{-1}%
  \endgroup
}

\title{Nonlinear analysis of a classical double oscillator model}

\author[1]{Bijan Bagchi}
\author[2]{Dibyendu Ghosh}
\author[3]{Lal Mohan Saha}

\affil[1]{Department of Physics, Shiv Nadar University, Gautam Buddha Nagar, Uttar Pradesh 201314, India} 
\affil[2]{Department of Mathematics, 
Barrackpore R. S. N. College,
West Bengal, Kolkata-700120, India}
\affil[3]{Department of  Mathematics, Shiv Nadar University, Gautam Buddha Nagar, Uttar Pradesh 201314, India} 

\date

\DeclareUnicodeCharacter{2212}{-}

\makeatletter
\renewcommand\@biblabel[1]{#1} 
\renewenvironment{thebibliography}[1]
     {\section*{\refname}%
      \@mkboth{\MakeUppercase\refname}{\MakeUppercase\refname}%
      \list{\@biblabel{\@arabic\c@enumiv}}%
           {\settowidth\labelwidth{\@biblabel{#1}}%
            \leftmargin\labelwidth
            \advance\leftmargin20pt
            \advance\leftmargin\labelsep
            \setlength\itemindent{-0.25pt}
            \@openbib@code
            \usecounter{enumiv}%
            \let\p@enumiv\@empty
            \renewcommand\theenumiv{\@arabic\c@enumiv}}%
      \sloppy
      \clubpenalty4000
      \@clubpenalty \clubpenalty
      \widowpenalty4000%
      \sfcode`\.\@m}
     {\def\@noitemerr
       {\@latex@warning{Empty `thebibliography' environment}}%
      \endlist}
\renewcommand\newblock{\hskip .11em\@plus.33em\@minus.07em}
\makeatother

\begin{document}
\maketitle
	
\begin{abstract}
    
\end{abstract}

\blfootnote{
{bbagchi123@gmail.com}}



Abstract
\maketitle
A classical double oscillator model, that includes in certain parameter limits, the standard harmonic oscillator and the inverse oscillator, is interpreted as a dynamical system. We study its essential features and make a qualitative analysis of orbits around the equilibrium points, period-doubling bifurcation, time series curves, surfaces of section and Poincar\'e maps. An interesting outcome of our findings is the emergence of chaotic behaviour when the system is confronted with a periodic force term like $f\cos \omega t$. \\

\section{Introduction}
\maketitle
Many years ago, Mathews and Lakshmanan \cite{Mat1} proposed a nonlinear deformation of the harmonic oscillator that is guided by a certain parameter $\lambda$ and which accounts for a non-polynomial oscillator term. They considered the Lagrangian to be given by

\begin{equation}
L = \frac{1}{2} (1 + \lambda x^2)^{-1} (\dot {x}^2 -\alpha^2 x^2), \quad -\frac{1}{\sqrt{|\lambda}|} < x < \frac{1}{\sqrt{|\lambda}|}
\end{equation}
where $\alpha$ is a non-negative real constant. Without $\lambda$ in $L$, the harmonic oscillator is recovered. 

The above Lagrangian also characterizes a mechanical system endowed with a position-dependent-mass (PDM) \cite{Mat2}. The subject of PDM \cite{Roos} has evinced a lot of interest in the literature and has developed into a separate branch of research over the past few decades \cite{Bag1, Dut, Mus, Dha, Car1, Sara, Fer, Dib}. In this connection, the relevance of PDM has been noteworthy in such areas as that of compositionally graded crystals \cite{Gel}, quantum dots \cite{Ser}, liquid crystals \cite{Bar} etc. and some classical problems possessing quantum analogs, for example, in branched Hamiltonian systems \cite{Bag2}.

Returning to the Lagrangian $L$, apart from the presence of a rational non-oscillator potential 

\begin{equation}
 V_{ML}(x) = \frac{1}{2} \frac{\alpha^2 x^2}{1+ \lambda x^2}
\end{equation}
it also supports a PDM defined by

\begin{equation}
 M(x) = \frac{1}{2} \frac{1}{1 + \lambda x^2}
\end{equation}
Evidently with the restriction of the configuration space specified in $(1)$ the mass function $M(x)$ is free from any singularity. Such a form for $M(x)$ has been studied in different contexts \cite{Yes}. One notices that $M(x)$ asymptotically approaches a zero-value with respect to $x$.

The equation of motion resulting from $(1)$ reads

\begin{equation}
(1+\lambda x^2)\ddot{x} - \lambda x \dot{x}^2 +\alpha^2 x =0
\end{equation}
which belongs to the quadratic Li\'enard class represented by the form 

\begin{equation}
\ddot{x} + r(x) \dot{x}^2 + s(x) =0
\end{equation}
where $r(x)$ and $s(x)$ stand for

\begin{equation}
r(x) = -\frac{\lambda x}{1+\lambda x^2}, \quad s(x) = \frac{\alpha^2 x}{1 +\lambda x^2}
\end{equation}

Among the properties of the above Lagrangian include the observation of 
Cari\~{n}ena et al \cite{Car2, Car3} that the kinetic term is invariant under the tangent lift of the vector field $X_x (\lambda) = \sqrt{1+\lambda x^2} \frac{\partial}{\partial x}$, the setting up of a quantum analogue in the form of a PDM problem and solving it for all possible orderings and including the effects of a linear dissipative term and demonstrating that in the presence of a two-parameter deformation \cite{Que} of the guiding potential, one runs into a chaotic behaviour of the whole system if the ranges of the coupling constants are suitably adjusted \cite{Bag3}. It may be remarked that in most of the investigations (see, for instance, \cite{nay, wig}) of dynamical systems, including seeking solitary wave solutions for standard systems \cite{pan}, the prototype of a second-order nonlinear autonomous differential equations is employed even for bifurcation analysis for complicated systems under localized perturbations \cite{zno}.

Recently, Schulze-Halberg and Wang \cite{Hal} have considered modifications to the Mathews-Lakshmanan Lagrangian $(1)$ by bringing in additional terms that not only account for the interactions present in $(1)$ but feature a one-parameter harmonic or
inverse harmonic oscillator contribution. Their two-parameter Lagrangian reads\footnote{In $L_{SW}$ we changed the sign of $\lambda$ and $\beta$ without loss of generality}

\begin{equation}
L_{SW} = \frac{1}{2} (1+ \lambda x^2)^{-1} [\dot{x}^2 -  (1 + \lambda)x^2] + \frac{\beta}{2} x^2, \quad \lambda \in \Re
\end{equation}
Here the mass-function is same as that in $(3)$ but the potential is modified to 

\begin{equation}
V_{SW} = \frac{1}{2} \frac{(1+ \lambda) x^2}{1+\lambda x^2} - \frac{\beta}{2} x^2, \quad  \beta \in \Re 
\end{equation}
which represents a double oscillator expressed as a combination of the harmonic oscillator $(\beta = - 1, \lambda \rightarrow 0)$ and the inverted harmonic oscillator $(\beta = 3, \lambda \rightarrow 0)$. Mathews-Lakshamanan nonlinear oscillator is recovered when $\beta = 0$. Specifically,  when $\beta = 0$ and $\alpha^2$ is set equal to $1+\lambda$ i.e. $\alpha^2 = 1+ \lambda$, one gets back the Lagrangian $(1)$ which is well known to be exactly solvable. However this is not the case when $\alpha^2 \neq 1+\lambda$ and $\beta \neq 0$. The classical dynamics for the system has been studied in detail in \cite{Hal} in connection with the existence of series solutions along with seeking suitable quantum analogs. Another paper by Blum and Elze \cite{blum} also investigated a double-well system semi-classically and pointed out the tendency of a chaotic behaviour.  In this paper we pursue a different line of approach in that although we use the same Lagrangian as given by $(7)$, we treat the equation of motion resulting from it as a coupled pair to look upon them from the perspective of a dynamical system \cite{Mat2, Bag4}. Subsequently we perform a qualitative analysis for the determination of local properties following from them and then include an
external periodic forcing term to study its influence as an additional damping effect. As we show explicitly we run into a chaotic behavior apart from obtaining period doubling bifurcation, time series curves, surfaces of section and Poincar\'e map. Estimates of Lyaunov exponents are also made. 

\section{Existence of fixed points and a qualitative analysis}

The equation of motion following from $L_{SW}$ reads\footnote{The second term in the equation of motion in \cite{Hal} has been corrected.}

\begin{equation}
(1+\lambda x^2) \ddot{x} - \lambda x \dot{x}^2 + (1-\beta + \lambda) x - 2\beta \lambda x^3 - \beta \lambda^2x^5 = 0
\end{equation}
It is of quintic type in the presence of a dissipative term which is present in the Mathews-Lakshmanan model as well. 

By introducing an auxiliary variable $y$ let us re-write $(9)$ in the form

\begin{eqnarray}
&& \dot{x} = y \nonumber \\
&&  \dot{y} = \frac{\lambda x y^2}{1+\lambda x^2} - (1- \beta + \lambda)\frac{x}{1+\lambda x^2} + 2\beta \lambda \frac{x^3}{1+\lambda x^2} + \beta \lambda^2 \frac{x^5}{1+ \lambda x^2}
\end{eqnarray}

The solutions of $\dot{x} = 0$ and $\dot{y} = 0$ furnish the fixed points. These are easily seen to be located on the $x$-axis with the squares of the abscissa satisfying

\begin{equation}
x^2_{1,2} = \frac{-\beta \pm [(1 + \lambda)\beta]^\frac{1}{2}}{\beta \lambda}, \quad \beta, \lambda \neq 0
\end{equation}
The positivity of the discriminant leads to the restriction that $(1+\lambda)\beta > 0$. As equilibrium points are real, the sign of $(1+\lambda)$ and that of $\beta$ has to be the same. Taking both to be positive, one obtains

\begin{equation}
    \lambda > - 1, \quad \beta \geq 0
\end{equation}
Specifically, taking $\beta = \frac{1}{2}$ and  $\lambda = - \frac{1}{2}$, the three equilibrium points are found to be placed at

\begin{equation}
     P_0^* = (0, 0), \quad P_1^* = (2, 0), \quad P_3* = (-2, 0)
\end{equation}

We now proceed to write down the Jacobian matrix which is to be evaluated at the critical points $x_c$

\begin{eqnarray}
 J &=& \left(
          \begin{array}{cc}
           0  & 1  \\
            J_{21} & 0  \\
          \end{array}
        \right)
   \end{eqnarray}
Note that is is off-diagonal with the expression for the element $J_{21}$ being

\begin{equation}
J_{21} = \frac{-(1-\beta + \lambda) + 6\beta \lambda x_c^2 +5\beta \lambda^2 x_c^4}{1+\lambda x_c^2} + \frac{2\lambda (1-\beta +\lambda)x_c^2 -4\beta\lambda^2 x_c^4 -2\beta\lambda^3 x_c^6}{(1+\lambda x_c^2)^2}
\end{equation} 
where $x_c$ corresponds to the solutions $x_{1,2}$ of $(11)$. Further $J_{21}$ is an even function in $x_c$. Among the equilibrium points $P_0^* = (0, 0)$ is neutrally stable but both $P_1^*$ and $P_2^*$ are stable and of elliptic type. 
Orbits and phase portrait about the equilibrium point $(2, 0)$ are presented in Figure 1.

Before we embark on the inclusion of a periodic force term in $(9)$, it is interesting to inquire into the possibility of what happens if the parameters $\beta$ and $\lambda$ satisfed the constraint 

\begin{equation}
    1-\beta + \lambda =0
\end{equation}
 In such a case a great simplification occurs with the fixed points $x_{1,2}^2$ emerging as $x_{1,2}^2 = \frac{2}{1-\beta}$ with $J_{21}$ acquiring the form

\begin{equation}
J_{21} = \frac{\beta \lambda x^2 (6+5\lambda x_c^2)}{1+\lambda x_c^2} -  \frac{2\beta \lambda^2 x_c^4 (2+\lambda x_c^2)}{(1+\lambda x_c^2)^2}
\end{equation}
With a trial value of $\beta = \frac{1}{2}$ or equivalently $\lambda = -\frac{1}{2}$, plausible solutions of $x_c$ are $\pm 2$. In both the cases $J_{21} = -4$. It signals that the eigenvalues of the matrix $J$ correspond to a pair of imaginary quantities. 
A linear stability analysis for different choices of the parameter values $\beta$ and $\lambda$ reveals, that for the reality of fixed points, one has to conform to either the conditions $1+\lambda > 0$ and  $\beta > 0$ or the pair $1+\lambda < 0$ and $\beta < 0$. Taking the former case without any loss of generality and choosing for illustration the numerical values  $\beta = 0.5$ and $\lambda = − 0.5$, the fixed points turn out to be $(0, 0), (2, 0)$ and $(− 2, 0)$. Thus the stability analysis points to the following results:\\

(i)	The point $(0, 0)$ is neutrally stable as for the case $(12)$.\\

(ii) Further, similar to the previous case, the fixed points $(2, 0)$ and $(− 2, 0)$ point to the eigenvalues $\pm 2i$ and imply that these are of elliptic (or centre) type. Hence the fixed points may be identified as stable points.\\




\section{Inclusion of a periodic forcing term}

The equation of motion when a periodic forcing term $f\cos \omega t$ is included is of the form 

\begin{equation}
(1+\lambda x^2) \ddot{x} - \lambda x \dot{x}^2 + (1-\beta + \lambda) x - 2\beta \lambda x^3 - \beta \lambda^2x^5 = f\cos \omega t
\end{equation}

The above nonautonomous systems can be re-expressed in terms of a set of three
autonomous nonlinear systems as arranged below

\begin{eqnarray}
&& \dot{x} = y \nonumber \\
&&  \dot{y} = \frac{\lambda x y^2}{1+\lambda x^2} - (1- \beta + \lambda)\frac{x}{1+\lambda x^2} + 2\beta \lambda \frac{x^3}{1+\lambda x^2} + \beta \lambda^2 \frac{x^5}{1+ \lambda x^2} + \frac{f \cos z}{1 + \lambda x^2} \nonumber\\
&& \dot{z} = \omega
\end{eqnarray}
In the following section the numerical simulation of the above scheme is carried out.

\section{Simulation results}

The behavior of the coupled system $(19)$ changes drastically when the periodic forcing term is included. In particular, the character of bifurcation phenomena is profoundly changed. Keeping parameters value  $\beta = 0.15, \lambda = -0.1, \omega = 1$ fixed and varying forcing amplitude $f$, the period doubling phenomena (Figure 2), following chaos, is observed for the system (2.1), as shown in Figure 3.

To observe the influence of the forcing term $f\cos{\omega t}$ in equation $(18)$, we keep in mind the fact that the Poincar\'e map serves as the basic tool for the understanding of the stability and bifurcations of periodic orbits. The time series curves, surfaces of sections as well as the Poincar\'e maps are drawn by keeping fixed the values of the parameters at $\beta = 0.5, \lambda = − 0.5, \omega = 1$ and different increasing values of amplitude $f$ as shown through various plots in Figure 4. One notices the transformation of regular periodic evolution to a chaotic evolution as the amplitude is increased from the value 0.1 to 0.71. To be specific, orbits start to disintegrate around the value of f = 0.55 when the motion is of quasi-periodic type. However, at the values $\beta = 0.5, \lambda =0.01, \omega =1$ and $f = 0.65, f = 0.71$, the motion turns completely chaotic. Surfaces of sections and Poincar\'e maps at each instant are significant and interesting. 
The plots of Lyapunov exponents for these two cases are given in Figure 5.\\


\section{Summary}

We have investigated in this paper a classical double oscillator model that includes in certain parameter limits, the standard harmonic oscillator and the inverse oscillator. By interpreting it as a dynamical system we have made a qualitative analysis of orbits around the equilibrium points, period-doubling bifurcation, time series curves, surfaces of section and Poincar\'e maps. The main highlight of our results is that a chaotic behaviour is noticed when a periodic force term like $f\cos \omega t$ is imposed upon the system. 

\section{Acknowledgments}

We would like to thank Rahul Ghosh for his help in organizing the figures.

\section{Data Avaialability}

Data sharing is not applicable to this article as no new data were created or analyzed in this study.

\newpage









\newpage

  \includegraphics[width=1\linewidth]{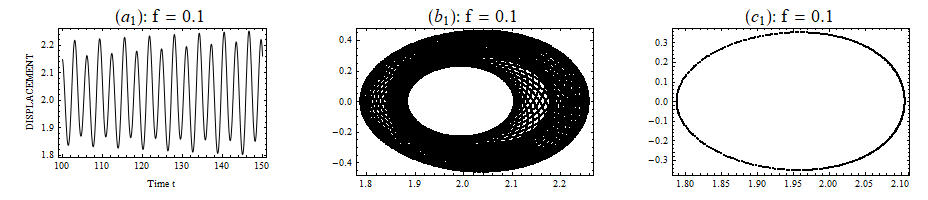}
 \begin{figure}
\centering
\begin{subfigure}{.5\textwidth}
  \centering
  \includegraphics[width=1\linewidth]{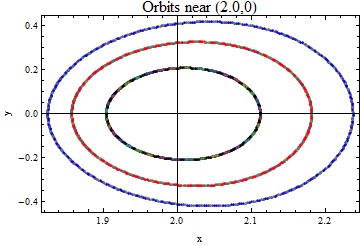}
  \caption{}
  \label{fig:2.1}
\end{subfigure}%
\begin{subfigure}{.4\textwidth}
  \centering
  \includegraphics[width=1\linewidth]{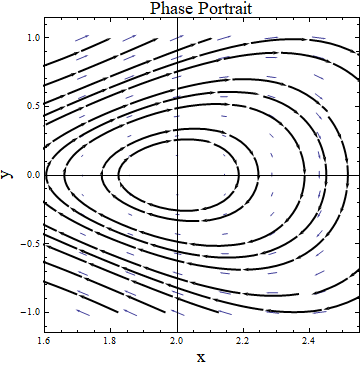}
  \caption{}
  \label{fig:2.2}
\end{subfigure}%
\caption{Orbits around the equilibrium (2, 0) and flow there shown, respectively, in left and right plots }
\label{fig:2}
\end{figure}
\begin{figure}
\centering
\begin{subfigure}{1\textwidth}
  \centering
  \includegraphics[width=1\linewidth]{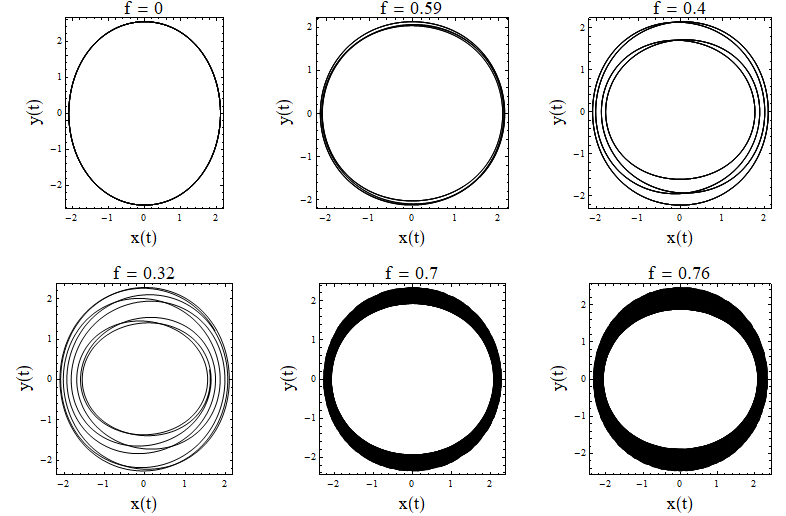}
  \caption{}
\end{subfigure}%
\caption{Period doubling bifurcation followed by chaos for system (2.1). Values of other parameters are $\beta = 0.15$, $\lambda = -0.1$, $\omega = 1$. }
\label{fig:3}
\end{figure}

\begin{figure}
\centering
\begin{subfigure}{1\textwidth}
  \centering
  \label{fig:4.1}
\end{subfigure} \\
\begin{subfigure}{1\textwidth}
  \centering
  \includegraphics[width=1\linewidth]{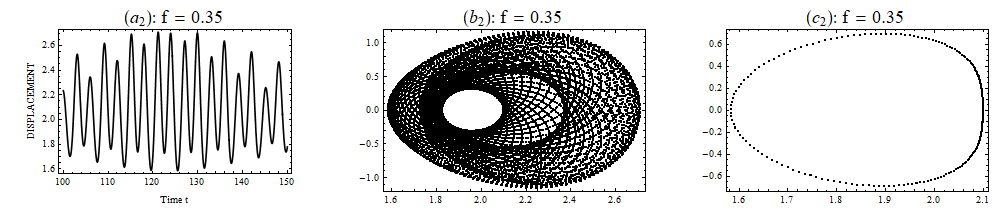}
  \caption{}
  \label{fig:4.2}
\end{subfigure} \\
\begin{subfigure}{1\textwidth}
  \centering
  \includegraphics[width=1\linewidth]{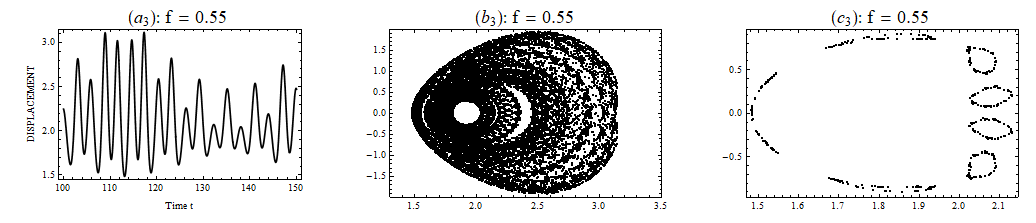}
  \caption{}
  \label{fig:4.3}
\end{subfigure} \\
\begin{subfigure}{1\textwidth}
  \centering
  \includegraphics[width=1\linewidth]{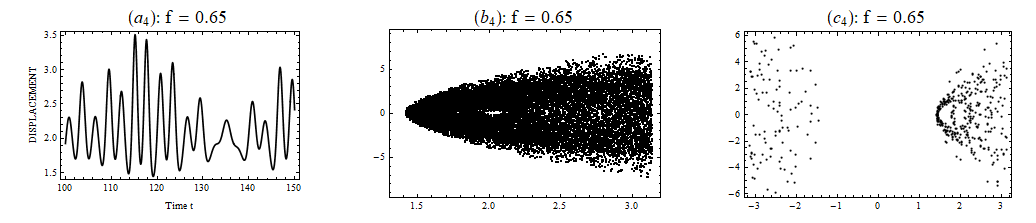}
  \caption{}
  \label{fig:4.4}
\end{subfigure} \\
\begin{subfigure}{1\textwidth}
  \centering
  \includegraphics[width=1\linewidth]{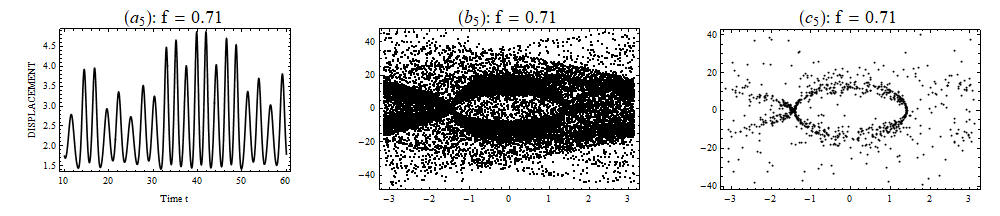}
  \caption{}
  \label{fig:4.5}
\end{subfigure}
\caption{ Plots of (i) Left column: Time series curves, (ii) Middle column: surfaces of section and (iii) Right Column: Poincaré maps, of system (2.1). }
\label{fig:4}
\end{figure}

\begin{figure}
\centering
\begin{subfigure}{1\textwidth}
  \centering
  \includegraphics[width=.8\linewidth]{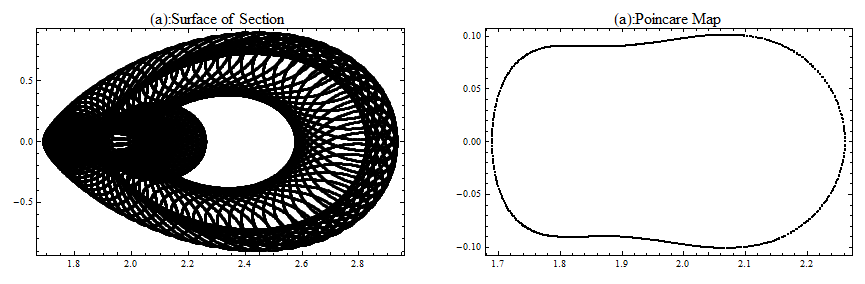}
  \caption{}
  \label{fig:5.1}
\end{subfigure} \\
\begin{subfigure}{1\textwidth}
  \centering
  \includegraphics[width=.8\linewidth]{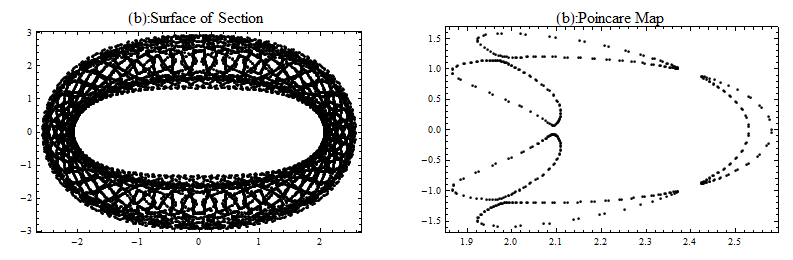}
  \caption{}
  \label{fig:5.2}
\end{subfigure} \\
\begin{subfigure}{1\textwidth}
  \centering
  \includegraphics[width=.8\linewidth]{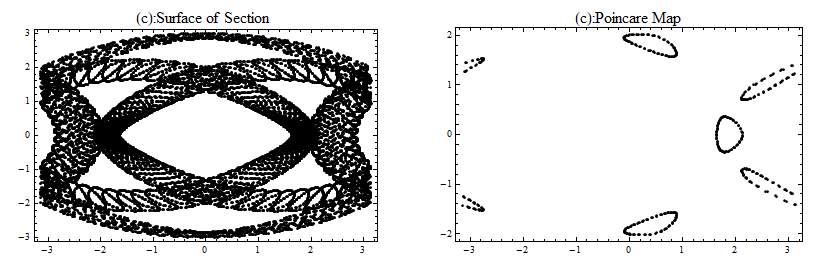}
  \caption{}
  \label{fig:5.3}
\end{subfigure} \\
\begin{subfigure}{1\textwidth}
  \centering
  \includegraphics[width=.8\linewidth]{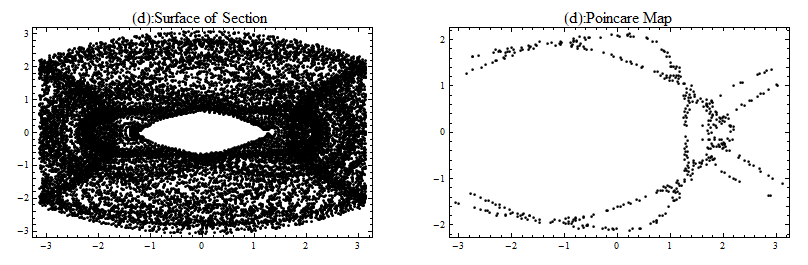}
  \caption{}
  \label{fig:5.4}
\end{subfigure} \\
\begin{subfigure}{1\textwidth}
  \centering
  \includegraphics[width=.8\linewidth]{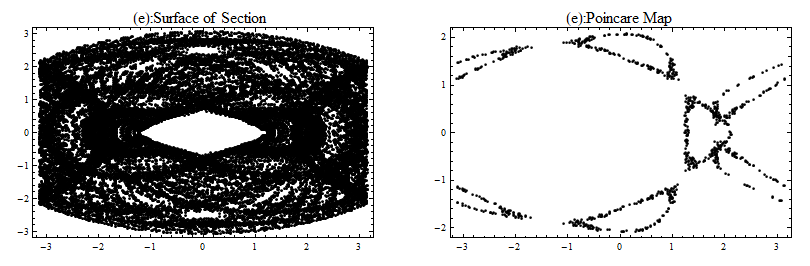}
  \caption{}
  \label{fig:5.5}
\end{subfigure}
\caption{Plots of surface of section and Poincaré map at different parameter spaces. For all cases, $\omega =1$ and (a) $\beta = 0.25$, $\lambda = -0.5$, $f = 0.65$, (b) $\beta = 0.5$, $\beta = -0.1$, $ f = 0.7$, (c) $\beta = 0.5$, $\lambda = 0.01$, $f = 0.7$ (d) $\beta = 0.5$, $\lambda = 0.01$, $f = 0.75$, (e) $\beta = 0.5$, $\lambda = 0.01$, $f = 0.76$. }
\label{fig:5}
\end{figure}

\begin{figure}
\centering
\begin{subfigure}{.5\textwidth}
  \centering
  \includegraphics[width=1\linewidth]{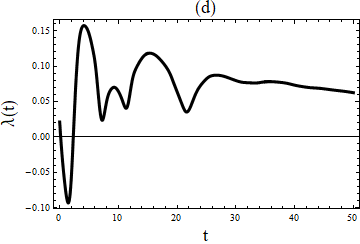}
  \caption{}
  \label{fig:7.1}
\end{subfigure}%
\begin{subfigure}{.5\textwidth}
  \centering
  \includegraphics[width=1\linewidth]{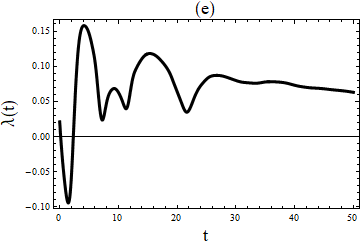}
  \caption{}
  \label{fig:7.2}
\end{subfigure}
\caption{Plots of Lyapunov exponents for two chaotic cases.}
\label{fig:7}
\end{figure}


\newpage 

\begin{thebibliography}{35}

\bibitem{Mat1}  P. M. Mathews and M. Lakshmanan, On a unique nonlinear oscillator, \textit{Quarterly Journal of Applied Mathematics} \textbf{32}, 215 (1974).

\bibitem{Mat2} S. Karthiga, V. Chithiika Ruby, M. Senthilvelan and  M. Lakshmanan, Quantum solvability of a general ordered position dependent mass system: Mathews-Lakshmanan oscillator, \textit{J. Math. Phys.} \textbf{58}, 102110 (2017).

\bibitem{Roos} O. von Roos, Position-dependent effective masses in semiconductor theory, \textit{Phys. Rev.} \textbf{B 27}, 7547 (1983).

\bibitem{Bag1} B. Bagchi, P. S. Gorain, C. Quesne, and R. Roychoudhury, A general scheme for the effective-mass Schr\"odinger equation and the generation of the associated potentials, \textit{Mod. Phys. Lett.} \textbf{A 19}, 2765 (2004).

\bibitem{Dut} A. de Souza Dutra, and C. A. S.  Almeida, Exact solvability of potentials with spatially dependent effective masses, \textit{Phys. Lett.}  \textbf{A 275}, 25 (2000).

\bibitem{Mus} O. Mustafa and S. H. Mazharimousavi, 
Quantum particles trapped in a position-dependent mass barrier; a d-dimensional recipe, \textit{Phys. Lett.} \textbf{A 358}, 259 (2006).

\bibitem{Dha} A. Dhahbi, Y. Chargui1 and A. Trablesi, A new class of exactly solvable models within the  Schr\'{o}dinger equation with position dependent mass, \textit{J. App. Maths. Phys.} \textbf{7}, 1013 (2019).

\bibitem{Car1} J. F. Cari\~{n}ena, J. F., M. F. Ra\~{n}ada and  M. Santander, The quantum harmonic oscillator on the sphere and the hyperbolic plane, \textit{Ann Phys.} \textbf{322}, 2249 (2007).

\bibitem{Sara} S. Cruz y Cruz and O. Rosas-Ortiz, Dynamical Equations, invariants and spectrum generating algebras of mechanical systems with position-dependent mass, \textit{SIGMA} \textbf{9}, 004 (2013).

\bibitem{Fer} M. I. Estrada-Delgado and D. J. Fern\'{a}ndez C, Ladder operators for the Ben Daniel-Duke Hamiltonians and their SUSY partners, \textit{Eur. Phys. J. Plus} \textbf{134}, 341 (2019).

\bibitem{Dib} D. Ghosh and B. Roy, Nonlinear dynamics of classical counterpart of the generalized quantum nonlinear oscillator driven by position dependent mass, \textit{Ann. Phys.} \textbf{134}, 353 (2015).

\bibitem{Gel} M. R. Geller and W. Kohn, Quantum mechanics of electrons in crystals with graded composition, \textit{Phys. Rev. Lett.} \textbf{70}, 3103 (1993).

\bibitem{Ser} L. Serra and E. Lipparini, Spin response of unpolarized quantum dots, \textit{Europhys. Lett.} \textbf{40}, 667 (1997).

\bibitem{Bar} M. Barranco, M. Pi, S. M. Gatica, E. S.  Hern\'{a}ndez, and J. Navarro, Structure and energetics of mixed Helium-4 - Helium-3 drops, \textit{Phys. Rev.} \textbf{B 56}, 8997 (1997).

\bibitem{Bag2} B. Bagchi, D. Ghosh and T. R. Tummuru, Branched Hamiltonians for a quadratic type Li\'enard oscillator, \textit{J. Non. Evol. Eqns and Appl.} \textbf{2018}, 101 (2020).

\bibitem{Yes} \"{O}. Ye\c{s}ilta\c{s}, Position-dependent mass approach and quantization for a torus Lagrangian, \textit{Eur. Phys. J. Plus} \textbf{131}, 308 (2016).

\bibitem{Car2} J. F. Cari\~{n}ena, M. F. Ra\~{n}ada and M. Santander, Two important examples of nonlinear oscillators \textit{Procs. of the 10th Int. Conf. in Modern Group Analysis Larnaca, Chipre},
edited by N. H. Ibragimov, Ch. Sophocleous and P. A. Damianou, (University of Cyprus) 39 (2005).

\bibitem{Car3} J. F. Cari\~{n}ena, M. F. Ra\~{n}ada and M. Santander, One-dimensional model of a quantum nonlinear harmonic oscillator, \textit{Rep. Math.
Phys.} \textbf{54}, 285 (2005).

\bibitem{Que} C. Quesne, Generalized nonlinear oscillators with quasi-harmonic behaviour: Classical solutions, \textit{J. Math. Phys.} \textbf{56}, 012903 (2015).


\bibitem{Bag3} B. Bagchi, S. Ghosh, S., B. Pal and S. Poria,  Qualitative analysis of certain generalized classes of quadratic oscillator systems, \textit{J. Math. Phys.} \textbf{57}, 022701 (2016).

\bibitem{nay} A. H. Nayfeh and D. T. Mook,  \textit{Nonlinear oscillations}, John Wiley Sons, New York (1995).

\bibitem{wig} S.  Wiggins, \textit{Introduction to Applied Nonlinear Dynamical Systems and Chaos}, SpringerVerlag, New York (2003).

\bibitem{pan} T. S. Raju, C. N. Kumar and P. K. Panigrahi, On exact solitary wave solutions of nonlinear Schr\"odinger equation with source, \textit{J Phys A: Math. Gen.} \textbf{38}, L271 (2005).

\bibitem{zno} D. I. Borisov, D. A. Zezyulin and M. Znojil,  Bifurcations of thresholds in essential spectra of elliptic
operators under localized non-Hermitian perturbations, \textit{Stud Appl Math.} \textbf{146}, 834 (2021).

\bibitem{Hal} A. Schulze-Halberg and J. Wang, Two-parameter double-oscillator model of Mathews-Lakshmanan type: Series solutions and supersymmetric partners, \textit{J. Math. Phys.} \textbf{56}, 072106 (2015).

\bibitem{blum} T. Blum and H.-Th Elze, Semiquantum Chaos in the Double-Well, \textit{Phys. Rev.}  \textbf{E 53}, 3123 (1996).

\bibitem{Bag4} B. Bagchi, \textit{Advanced Classical Mechanics}, Taylor and Francis (2017).





































\end{thebibliography}
\end{document}